\documentclass{aa}
\usepackage{graphics}
\usepackage{txfonts}
\usepackage{natbib}
\bibpunct{(}{)}{;}{a}{}{,}

\DeclareRobustCommand{\ion}[2]{%
\relax\ifmmode
\ifx\testbx\f@series
{\mathbf{#1\,\mathsc{#2}}}\else
{\mathrm{#1\,\mathsc{#2}}}\fi
\else\textup{#1\,{\mdseries\textsc{#2}}}%
\fi}

\def\apjl{ApJ}%
\def\apjs{ApJS}%
\def\ao{Appl.~Opt.}%
\def\aap{A\&A}%
\begin{document}
\title{Relation between photospheric magnetic field and chromospheric emission}

\author{R. Rezaei\inst{1}, R. Schlichenmaier\inst{1}, C.A.R. Beck\inst{1,2}, J.H.M.J. Bruls\inst{1}, 
and W. Schmidt\inst{1}}

\institute{Kiepenheuer-Institut f\"ur Sonnenphysik, Sch\"oneckstr. 6, 79\,104 Freiburg, Germany
\and
Instituto de Astrof\'isica de Canarias~(IAC), E 38\,205, La Laguna, Espain}
\date{Received 22 December 2006/Accepted 25 January 2007}
\titlerunning{Relation between  photospheric magnetic field and chromospheric emission}
\authorrunning{Rezaei et al.}
\offprints{rrezaei@kis.uni-freiburg.de}
\abstract
{} 
{ We investigate the relationship between the photospheric magnetic field and the 
emission of the mid chromosphere of the Sun.} 
{We simultaneously observed the Stokes parameters of the photospheric iron line 
pair at 630.2\,nm and the intensity profile of the chromospheric \ion{Ca}{ii}\,H line at 396.8\,nm 
 in a quiet Sun region at a heliocentric angle of 53\degr. 
Various line parameters have been deduced from the \ion{Ca}{ii}\,H line profile. 
The photospheric magnetic field vector has been reconstructed from an inversion of the measured 
Stokes profiles.
After alignment of the Ca and Fe maps, 
a common mask has been created to define network and inter-network regions. 
We perform a statistical analysis of network and inter-network properties. 
The H-index is the integrated emission in a 0.1\,nm band around the Ca core.
We separate a non-magnetically, H$_{\textrm{non}}$, and a magnetically, H$_{\textrm{mag}}$, heated component 
from a non-heated component, H$_{\textrm{co}}$ in the H-index.}
{The average network and inter-network H-indices are equal to 12\,and\,10\,pm, respectively. 
The emission in the network is correlated with the magnetic flux density, 
approaching a value of $H$\,$\approx$\,10\,pm for vanishing flux. 
The inter-network magnetic field is dominated by weak field strengths with values down to 200\,G and
has a mean absolute flux density of about 11\,Mx\,cm$^{-2}$. }
{We find that  a dominant fraction of the calcium emission caused by the  heated atmosphere in the magnetic {network} 
has non-magnetic origin\,(H$_{\textrm{mag}}\approx2\,$pm, H$_{\textrm{non}}\approx3\,$pm). 
Considering the effect of straylight, 
the contribution 
from an atmosphere with no temperature rise to the H-index (H$_{\textrm{co}}\approx6\,$pm)
is about half of the observed H-index in the inter-network.   
The H-index in the inter-network is not correlated to any property 
of the photospheric magnetic field, suggesting that  magnetic 
flux concentrations have a negligible 
role in the chromospheric heating in this region. The height range of the thermal coupling 
between the photosphere and low/mid chromosphere increases in presence of magnetic field. 
In addition, we demonstrate that a poor signal-to-noise level in the Stokes profiles 
leads to a significant over-estimation of the magnetic field strength. }
\keywords{Sun: photosphere -- Sun: chromosphere -- Sun: magnetic fields}

\maketitle

\section{Introduction}
The dominant pattern covering the entire solar surface, except sunspots, is granulation, 
the top of small-scale convection cells with diameters of 1-2\,Mm. 
On a larger scale of $\sim$\,20\,Mm \textit{supergranules} are observed 
which show the same pattern as granulation: a horizontal flow from the center towards the boundary of the cell. 
These large scale convection cells have a 
mean lifetime of 20\,h, much longer than the granular time-scale (some 10 minutes). 
The  {\rm\it chromospheric network}  forms at the boundary of the supergranular cells. 
The network is presumably formed by the long--time advection of magnetic flux to the 
boundaries of the supergranules~\citep{priest_etal_02, cattaneo_etal_03}. The interior of the network cells, 
the {\rm\it inter-network},  has much less magnetic field than the network~\citep[e.g.,][]{keller_etal_94}.
The magnetic and thermodynamic properties of the network and inter-network are different~\citep[e.g.,][]{lites_02}.

The chromospheric heating mechanism is one of the main challenges of solar 
physics~\citep[][and references therein]{nar_ulm_96}. 
The core emission of the \ion{Ca}{ii}\,H and K lines is an important source of radiative losses 
in the chromosphere. 
Moreover, this emission is an important tool to study the temperature stratification 
and the magnetic activity of the outer atmosphere of the Sun and other stars~\citep[e.g.,][]{schrij_zwaan}. 
Most of the observational studies  based on these lines 
use either the calcium intensity 
profile~\citep[e.g.,][]{cram_dame_83, lites_etal_93} or combinations of 
calcium filtergrams and magnetograms~\cite[e.g.,][]{berger_title_01}. 
From the observation of the calcium spectrum alone, 
it is not possible to distinguish between the magnetic and non-magnetic heating components. 
Combining calcium filtergrams with magnetograms allows to separate those components, 
but the spectral information is lost.  
Simultaneous observations that allow to reconstruct the magnetic field and record the spectrum for the \ion{Ca}{ii}\,H line are
rare~\citep[e.g.,][]{lites_etal_99} and only available at lower spatial resolution. 
Therefore, it is not surprising that none of the present theories, 
mechanical and Joule heating, was confirmed or rejected observationally~\citep{fossum_carl, socas_n_05}.

\cite{schrijver_XI_87} studied a sample of late-type stars and 
introduced the concept of the {\em basal flux} to separate 
the 
non-magnetic heating from the magnetic one. 
The Sun is not a very active star, with a chromospheric radiative loss on the order of the 
basal flux for  Sun--like stars~\citep{fawzy_etal_02b}. 
\cite{carl_stein_97} presented a semi--empirical 
hydrodynamic model where enhanced chromospheric emission 
is due to outward propagating acoustic waves. 
This model was criticized in subsequent investigations~\citep[e.g.,][]{kalkofen_etal_99, fossum_carl},  
which was partly due to disagreements on the temperature stratification in 
higher layers~\citep[e.g.,][]{ayres_02, wedemeyer_etal_05}. 
In addition, it is not clear whether high- or low-frequency acoustic waves play the dominant 
role in the energy transport to higher layers~\citep{fawzy_etal_02a, jefferies_etal_06}. 
While there are theoretical indications that the magnetic filling factor  
discriminates between different regimes of heating ~\citep{sol_stein90}, there 
is no canonic model reproducing  thermal, dynamical and 
magnetic properties of the solar chromosphere~\citep{judge_peter_98, rutten_99}. 

The POlarimetric LIttrow Spectrograph \citep[POLIS,][]{schmidt03,beck05b} was designed to 
provide co--temporal and co--spatial measurements of the magnetic field in the photosphere 
and the \ion{Ca}{ii}\,H intensity profile. 
We use POLIS to address the question of the chromospheric heating mechanism,  
by comparing properties of network and inter-network in photosphere and chromosphere 
by a statistical analysis. 
With the information on the photospheric fields, 
we separate the contributions of the magnetically and non-magnetically heated component. 
For the first time we study the 
correlation of the chromospheric emission  with the corresponding  
amplitude/area asymmetry and Stokes-$V$ zero--crossing velocity at the corresponding photospheric position. 
We also investigate the magnetic field strength distribution of the inter-network to check whether 
it consists of weak fields~\citep[e.g.,][]{faurobert_etal_01, collados_01} or 
kilo-Gauss fields~\citep[e.g.,][]{almeida_cerdena_kneer03,alm_emon_cat_03}. 

Observations and data reduction are 
explained in Sects. 2 and 3.  Histograms of the obtained parameters are presented in Sect. 4. 
Correlations between the chromospheric and photospheric parameters are addressed in Sect. 5. 
The heating contributions are elaborated in Sect. 6. 
Discussion and conclusions are presented in Sects. 7 and 8, respectively. Details of the 
magnetic field parameters in the inter-network are discussed in Appendix A. 
The straylight contamination and calibration uncertainties for the POLIS Ca channel 
are estimated in Appendix B. An overview of some of our results also appears in~\cite{reza_etal_2}.

\section{Observations}
We observed a network area close to the active region NOAA 10675 on September 27, 2004, 
with the POLIS instrument at the German Vacuum Tower Telescope (VTT) in Tenerife. 
The observations were part of a coordinated observation campaign in the International 
Time Program, where also the Swedish Solar Telescope (SST), Dutch Open Telescope (DOT) 
and the T\'elescope H\'eliographique pour 
l'\'Etude du Magn\'etisme et des Instabilit\'es Solaires (TH\'EMIS) participated. 
Here we analyze a series of thirteen maps of the same network 
region at a heliocentric angle of 53\degr, taken during several hours. 
We achieved a spatial resolution  of around 1.0 arcsec, estimated from the 
spatial power spectrum of the intensity maps.

All maps were recorded with a slit width of 0.5 arcsec, 
a slit height of 47.5 arcsec, and an exposure time per slit position of 4.92 s. 
The scan extension for the first three maps was 40.5, 55.5, and 20.5 arcsec,  
while  it was 25.5 arcsec for the remaining ten maps.

The \ion{Ca}{ii}\,H line and the visible neutral iron lines at 630.15\,nm, 630.25\,nm, and 
\ion{Ti}{i} 630.38\,nm were observed with the blue (396.8\,nm) and red (630\,nm) channels of POLIS. 
The spatial sampling along the slit (y-axis in Fig.~1) was 0.29 arcsec. The scanning step was 0.5 arcsec for all maps. 
The spectral sampling of 1.92\,pm for the blue channel  
and 1.49\,pm for the red channel leads to a velocity dispersion of 
1.45 and 0.7\,km\,s$^{-1}$ per pixel, respectively. 
The spectropolarimetric data of the red channel 
were corrected for instrumental effects and telescope 
polarization with the procedures described by \cite{beck05b, beck05a}. 

\begin{figure*}
\resizebox{\hsize}{!}{\includegraphics{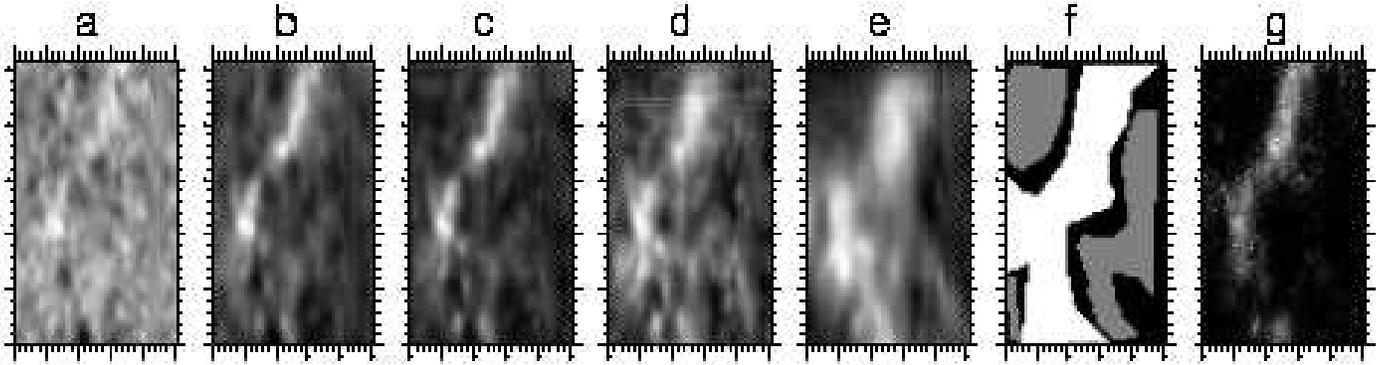}}
\caption[]{{\em From left to right:} a) the \ion{Fe}{i}\,630\,nm continuum intensity, 
b) the calcium wing intensity at 396.490\,nm which was calibrated to FTS 
data~\citep{stenflo_84}, c) the outer calcium wing intensity (W1), 
d) the inner calcium wing intensity (W3), e) the H-index, 
f) the masks which separate network (white) from the inter-network (gray), 
and g) the magnetic flux density obtained from the inversion. 
We did not use the black region between the network and inter-network. 
Each small tickmark  is 1\,arcsec. Note that sampling in x and y directions are different.}
\end{figure*}

\begin{table}
\begin{center}
\caption{The definition of the characteristic parameters of the \ion{Ca}{ii}\,H profile 
for the peak sample (upper part) and the band sample (lower part)~(see also Fig.~2). 
Wavelengths are in nm.}
\begin{tabular}{c c} 
\hline
quantity: peak sample & description  \\ \hline
H$_3$  &   core intensity\\
H$_{2\textrm{v}}$  &    violet emission peak\\
H$_{2\textrm{r}}$  &   red emission peak\\
V/R       &   H$_{2\textrm{v}}$/H$_{2\textrm{r}}$\\
emission strength &  H$_{2\textrm{v}}$/H$_3$ \\
$\lambda$(H$_3$)  &  calcium core wavelength\\
$\lambda$(H$_{2\textrm{v}}$)  &  H$_{2\textrm{v}}$ wavelength\\
$\lambda$(H$_{2\textrm{r}}$)  &  H$_{2\textrm{r}}$ wavelength\\\hline
quantity: band sample & description  \\ \hline
H-index  &   396.849\,$\pm$\,0.050\\ 
H$_3$     & 396.849\,$\pm$\,0.008 \\
H$_{2\textrm{v}}$ & 396.833\,$\pm$\,0.008  \\
H$_{2\textrm{r}}$ & 396.865\,$\pm$\,0.008 \\
W1  & outer wing: 396.632\,$\pm$\,0.005  \\
W2  & middle wing:  396.713\,$\pm$\,0.010   \\
W3  & inner wing:  396.774\,$\pm$\,0.010   \\
\hline
\end{tabular}
\end{center}
\end{table}

Figure 1 displays an overview of one of the thirteen maps after spatial alignment. 
The map  (a)  shows the \ion{Fe}{i}\,630\,nm continuum intensity normalized to 
the average quiet Sun intensity. 
The map (b) shows the \ion{Ca}{ii}\,H wing intensity, taken close 
to 396.490\,nm (for a definition of line parameters see Fig.~2 and Table 1). 
These two maps were used for the spatial alignment 
of the red and blue channels. 
The next two maps (c and d)  show the intensities in the outer and inner 
wings (W1, respectively, W3). 
The inner wing samples a wavelength band close to the core; hence, it is more influenced 
by the line--core emission and shows higher contrast of the network than the outer wing. 
The next map (e) is the H-index, i.e., the intensity of the calcium core integrated over 
0.1\,nm (cf.~Table 1). 
The network features appear broadest and show the highest contrast in this map. 
The map (f) demonstrates network and inter-network masks (Sect. 3.1).
The map (g) shows the magnetic flux density obtained from 
the spectro-polarimetric data (Sect. 3.2). 

\section{Data analysis}
In this section we discuss characteristic parameters of the calcium profile and the \ion{Fe}{i} 630\,nm line pair. 
We briefly explain the inversion method that was used to infer the vector magnetic field.

\subsection{Definition of network and inter-network}

For each map, we created a mask to distinguish between the network and inter-network regions (Fig.~1f). 
This was done manually on the basis of the magnetic flux and the H-index. 
We did not use the black region in Fig.~1f which separates network from inter-network. 
To study structures in  these two regions, we define two statistical samples:

\paragraph{\textbf{The peak sample:}} 
Each member of this sample has a regular $V$ profile and a double reversal in the calcium core. 
For this sample, we study quantities  related to
the \ion{Ca}{ii}\,H emission peaks~(Table 1, upper part). 
The number of data points in the network and inter-network 
are 19509  and 1712, respectively. 
We emphasize that the few inter-network points in this sample are not distributed uniformly. 
They usually belong to small-scale magnetic elements surrounded by large areas without  
magnetic signal above the noise level.

\paragraph{\textbf{The band sample:}} 
All network and inter-network points, marked in the masks, are present in this sample. 
Since some of these profiles do not have two emission peaks in the calcium core, 
we use integrated intensities in a fixed spectral range~\citep{cram_dame_83} at the 
core and wing of the calcium profile~(Table 1, lower part). 
This definition retrieves reasonable chromospheric parameters.  
In total from all the maps, 24225  points were selected as network (white in Fig.~1f) 
and 20855 as inter-network positions (gray in Fig.~1f).

\begin{figure}
\resizebox{\hsize}{!}{\includegraphics{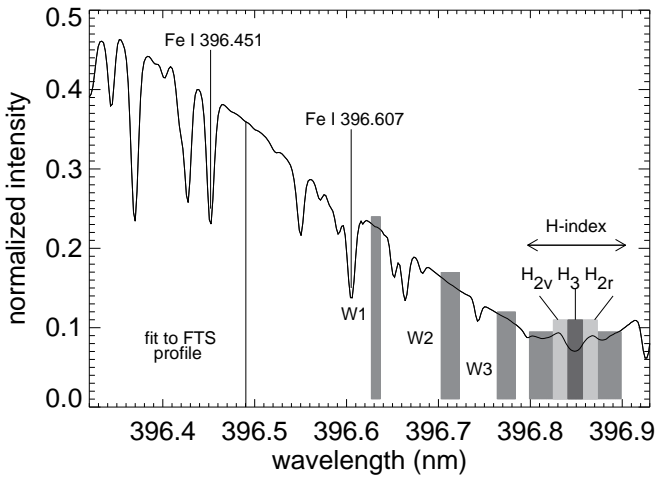}} 
\caption[]{Sample averaged calcium profile of one of the maps 
(the average profile is similar for all other maps). 
The bands are explained in Table 1.}
\end{figure}

\subsection{Analysis of \ion{Ca}{ii}\,H profiles}
We averaged profiles over a large area (including network) to obtain an average 
profile  for each map. These average profiles are the mean of more than three thousand profiles each time; 
an example is shown in Fig.~2. 
We then normalized the intensity  at the line wing at 396.490\,nm to the FTS profile~\citep{stenflo_84}. 
The same normalization coefficient was applied to all profiles of the map.  
Hence,  all profiles are normalized to  the average continuum intensity.

Table 1 lists the characteristic parameters we define for each calcium profile:
\begin{itemize}

\item[a)] The intensity of the \ion{Ca}{ii}\,H core, H$_3$, and the amplitudes of the emission 
peaks, H$_{2\textrm{v}}$ and  H$_{2\textrm{r}}$, and their respective 
positions, $\lambda_{\rm H_3}, \lambda_{\rm H_{2v}}$, 
and $\lambda_{\rm H_{2r}}$ (peak sample).

\item[b)] Intensities in the outer (W1), middle (W2), and inner line wings (W3), 
which are integrated over spectral bands with a \textit{fixed} spectral range (band sample).


\end{itemize}

From these quantities we derive: 
\begin{itemize}
\item[1.] the ratio of the emission peaks, V/R = H$_{2\textrm{v}}$/H$_{2\textrm{r}}$, and
\item[2.] the emission strength which is the intensity of the violet emission peak 
divided by the core, H$_{2\textrm{v}}$/H$_3$.
\end{itemize}

The band intensities are defined such that the wavelength intervals are fixed for all the 
profiles and do not depend on the peak/core positions (Table 1, lower part).  
Figure 3 shows a comparison between the H$_{2\textrm{v}}$ parameters derived from 
the band and peak definitions.
The correlation is very strong (also for other peak parameters). 
Thus, it is justified to use the values from the band sample to improve the statistics for the inter-network.

\begin{figure}
\resizebox{\hsize}{!}{\includegraphics*{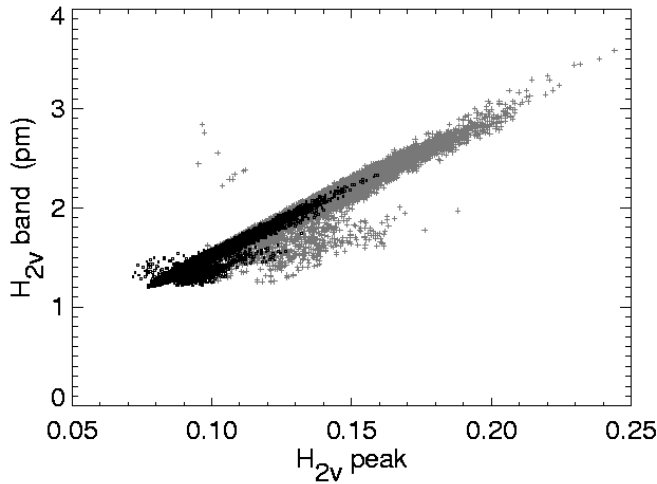}}
\caption[]{Scatter plot of the H$_{2\textrm{v}}$ band vs. peak definition. 
In the band definition, a fixed spectral range was used. 
Gray and black show the network and inter-network, respectively.}
\end{figure}

\begin{figure*}
\resizebox{16.3cm}{!}{\includegraphics{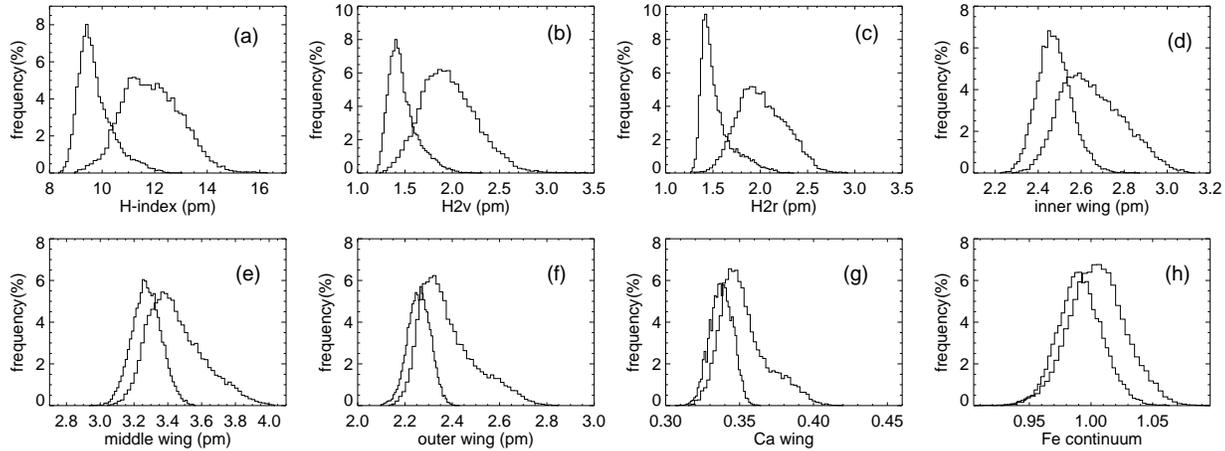}}
\caption[]{Histograms of the intensity parameters for the network (thick) and 
inter-network (thin), using the band sample: a) the H-index, b) H$_{2\textrm{v}}$, c) H$_{2\textrm{r}}$, 
d) W3\,(the inner wing intensity), e) W2\,(the middle wing intensity), 
f) W1\,(the outer wing intensity), g) the calcium wing intensity at 396.490\,nm which 
was used for normalization~(see Fig.~2), 
and h) \ion{Fe}{i}\,630.2\,nm normalized continuum intensity (see Table 1 for definitions). }
\end{figure*}

\subsection{Analysis of \ion{Fe}{i} 630\,nm line profiles}
The data of the red channel of POLIS were treated with the standard polarimetric calibration 
procedures described in \cite{beck05b, beck05a}. 
The spectral line curvature was removed using the telluric O$_2$ line at 630.20\,nm  with a routine 
described in \cite{reza_etal_1}. 
The polarization signal in $Q(\lambda)$, $U(\lambda)$, and $V(\lambda)$ is   
normalized to the local continuum intensity, $I_c$, for each pixel. 
The rms noise level of the Stokes parameters in the continuum  
was  $\sigma$\,=\,8.0\,$\times 10^{-4}$\,$I_c$.  
Only pixels with $V$ signals greater than 3\,$\sigma$  were 
included in the profile analysis. 
For regular Stokes--$V$ signals above the threshold, 
we derived positions and amplitudes of the profile extrema  
in all Stokes parameters  by fitting a parabola to each lobe. 

An inversion was performed using the SIR code \citep{sir92}. 
We used the same setup as in \cite{luis_beck05} and \cite{beck_etal_06}:
a two--component solar atmosphere model with one magnetic and one field-free component. 
Additionally, a variable amount of straylight was allowed for. 
This accounts for unresolved magnetic fields inside each pixel.
We did not consider any gradient for the atmospheric parameters except for the temperature. 
The inversion yields the magnetic field vector and an estimate for the magnetic filling fraction. 
The flux density map (magnetic flux per pixel, Fig.~1g) is 
based on the inversion results.

\section{Histograms}

\subsection{Parameters of the intensity profiles}
The network patches are brighter than the inter-network at all wavelengths: 
in the core and wing of the \ion{Ca}{ii}\,H line and in \ion{Fe}{i}\,630\,nm continuum. 
The distributions of the intensity parameters at different wavelengths are shown in Fig.~4. 
Thick and thin curves show network and inter-network, respectively. 
A comparison between the network and inter-network histograms indicates that 
the calcium core parameters (H-index, H$_{2\textrm{v}}$, H$_{2\textrm{r}}$) 
are more shifted than the wing/continuum intensities (Fig.~4a, b, c). 
In other words, the intensity contrast increases from the continuum toward 
the wing and peaks at the core. 
Moreover, the wing intensities show different behaviors 
due to their respective distances to the core: 
the closer the selected wavelength band to the core, the larger the shift between 
network and inter-network distributions.  
Hence, the histogram of 
the inner wing intensity (Fig.~4d) has a larger shift than the middle and outer  wings (Fig.~4e and f). 
The distribution of the  \ion{Ca}{ii}\,H wing\footnote{Here, wing means the wavelength 
that was used for the intensity calibration (see Fig.~2).} and \ion{Fe}{i}\,630\,nm continuum 
intensities also show a shift between the network and 
inter-network: the peak of the distribution in the network is slightly brighter than in the 
inter-network in 396.490\,nm and also in the 630\,nm continuum (Fig.~4g and h).

\begin{figure*}
\resizebox{16.3cm}{!}{\includegraphics{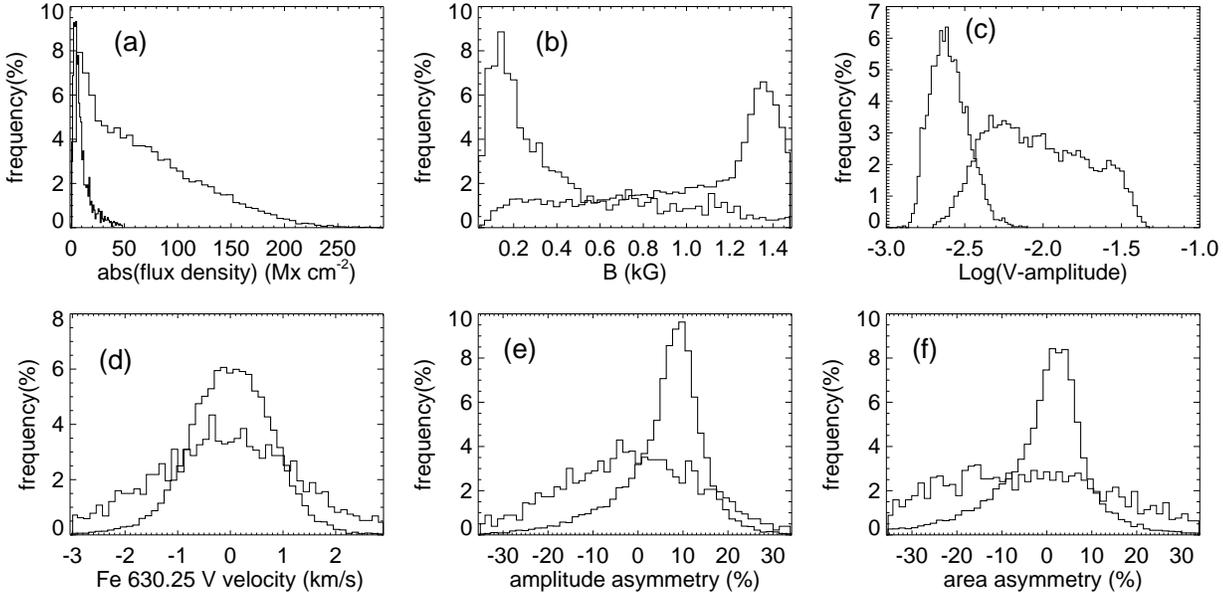}}
\caption[]{Histograms of the magnetic field parameters for the network (thick) and 
inter-network (thin), using the peak sample: a) the absolute magnetic flux density, 
b) the field strength, c) Stokes $V$ amplitude, d) \ion{Fe}{i}\,630.25\,nm $V$ velocity, 
e) the amplitude asymmetry, and f) the area asymmetry. }
\end{figure*}

\subsection{Magnetic field parameters}
From the inversion, we obtain the magnetic field and the  magnetic flux  
in the photosphere. The statistics of these quantities in the network and inter-network are 
discussed here; the relation with the chromospheric parameters is discussed  
in the next section. We restrict the analysis 
to those profiles with a regular Stokes--$V$ and a double emission calcium profile, 
although  there are only 1712 profiles in  the inter-network. 
The histogram of the absolute magnetic flux density, $\Phi\,=\,f\,B\,\cos\alpha$, is shown in Fig.~5a,  
where $f$ is the magnetic filling factor, $B$ is the field strength, and $\alpha$ is the angle 
between the magnetic field orientation and the line-of-sight. 
The histogram for the inter-network has a narrow distribution, which peaks at some  
$\sim$\,4\,Mx\,cm$^{-2}$\, whereas there is a long tail 
 up to $\sim$\,250\,Mx\,cm$^{-2}$\, for the network. 
The distribution function of the magnetic field strength (Fig.~5b) has a peak around $\sim$\,1.3\,kG for the 
network elements and increases with decreasing field strength for the inter-network.
The peak at 200\,G is due to the detection limit of the polarization signal.  
The histogram of the $V$ amplitude (Fig.~5c) shows weak inter-network signals with a  
$V$ amplitude far less than 1\%. 
The distribution of the inter-network $V$ amplitude\footnote{  
Histogram of the $V$-amplitude of the inter-network (Fig.~5c)  
includes all data points with a clear $V$ signal.} 
peaks at $\sim\,2\,\times\,10^{-3}\,I_c$ 
whereas the network shows a broad distribution up to $\sim\,0.03\,I_c$.

The $V$ velocity distribution is shown in Fig.~5d. Although it has a peak around zero both for the 
network and inter-network, the inter-network shows a larger fraction of high-velocity $V$ profiles. 
The amplitude and area asymmetries in the network peak at small positive values of 10\,\% and 
 3\,\%, respectively (Fig.~5e and f).  
In contrast, there is no tendency to positive or negative values for the inter-network asymmetries. 
This is in agreement with other studies of quiet Sun magnetic fields~\citep{sigwarth99, sigwarth01, khomenko_etal_03}.

\begin{table}
\begin{center}
\caption{Correlation coefficients between the band intensities and the H-index in 
the network and inter-network using the band sample.}
\begin{tabular}{c c c}\hline 
parameter          &  network   & inter-network \\ \hline 
H$_3$              & 0.88 & 0.90 \\\hline
H$_{2\textrm{v}}$  & 0.92 & 0.94 \\\hline
H$_{2\textrm{r}}$  & 0.88 & 0.93 \\\hline
W3                 & 0.77 & 0.68 \\\hline
W2                 & 0.73 & 0.53 \\\hline
W1                 & 0.58 & 0.36 \\\hline
\end{tabular} 
\end{center}
\end{table}

\section{Correlations}

\subsection{Correlations between chromospheric quantities}
There is a strong correlation between the H-index, H$_3$, H$_{2\textrm{v}}$, H$_{2\textrm{r}}$, 
and the inner wing intensity (with a correlation coefficient $\ge$\,0.8, see Fig.~6).  
Therefore, we adopt the H-index as a proxy for the calcium core emission. 
It also has the advantage to be the calcium intensity parameter closest to the widely used  
filtergrams\footnote{Note that filtergrams, e.g., from DOT~\citep{wijn_etal_05} 
also cover parts of the line wing while we integrate 
over a band with a width of 
.1\,$\pm$\,0.001\,nm.}.

The (V/R) ratio is an indicator for the H$_3$ line--core position \citep{rutten_rev91, cram_dame_83}. 
There is a strong correlation between the (V/R) ratio 
and $\lambda$(H$_3$): the redshifted 
and blueshifted calcium profiles correspond to (V/R) ratios  
smaller and larger than one, respectively (Fig.~6, lower right panel)\footnote{ The calcium core position 
is expressed in velocity units for comparison. 
This does not mean that it corresponds to a Doppler shift.}. 
We find a similar correlation between the emission strength and the (V/R) ratio: 
the larger the (V/R) ratio, the larger the emission strength.
The lower left panel of Fig.~6 shows the correlation between the outer wing intensity and the H-index. 
While there is a correlation in the network (gray), there is no significant correlation in the inter-network (Table 2). 
We calculated the correlation coefficient between the H-index and all
intensities in the line wing to investigate this difference in more
detail (Table 2). In the network, there is a stronger correlation than
in the inter-network for all the wing bands considered. 
We return to this point in Sect. 7.

\subsection{Correlations between photospheric quantities}

There is a strong correlation between the 
amplitude and area asymmetries, both in the network and inter-network (Fig.~7, top left panel).  
In contrast, there is no correlation between either amplitude or area asymmetries and 
the $V$ velocity (bottom left panel, Fig.~7).   
These are  typical properties of the quiet Sun magnetic field \citep[e.g.,][]{sigwarth99}. 
The right panels of Fig.~7  show scatter plots of the $V$ asymmetries versus $V$ amplitude. 
There is a  tendency for high amplitude $V$ signals to have small asymmetries. 
On the other hand, 
histograms of the asymmetries peak at a positive value. 
Implications of these two findings will be discussed in Sect. 5.3.2.

\begin{figure}
\resizebox{\hsize}{!}{\includegraphics{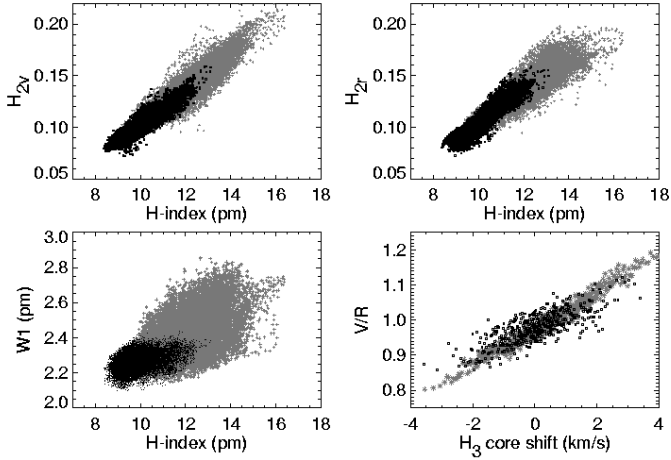}} 
\caption[]{{\em Upper panels}: correlation between the H$_{2\textrm{v}}$ and H$_{2\textrm{r}}$ based on the 
peak sample and the H-index.  
{\em Lower left}: correlation between W$_1$ and the H-index. 
{\em Lower right}:  correlation of the (V/R) ratio with the calcium core position. 
Gray and black show the network and inter-network, respectively (for abbreviations, see Table 1).}
\end{figure}

\begin{figure}
\resizebox{\hsize}{!}{\includegraphics{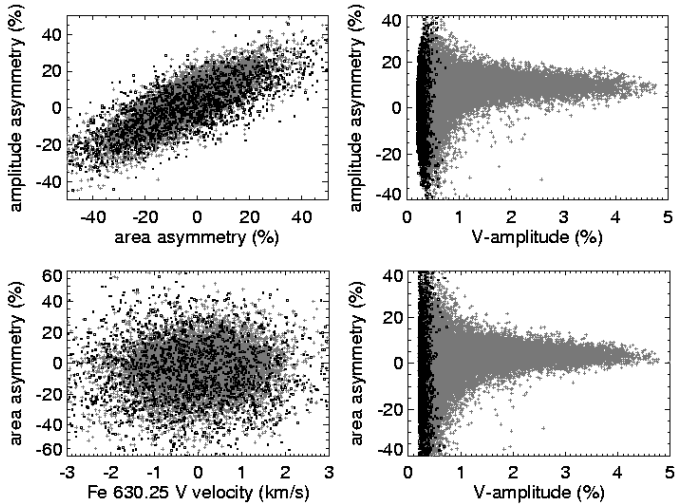}}
\caption[]{{\em Top left}: scatter plot of the amplitude vs. area asymmetries. 
{\em Bottom left}: scatter plot of the area asymmetry vs. the $V$ velocity. 
{\em Right panels}:  scatter plots of the amplitude and area asymmetries  vs. the \ion{Fe}{i}\,630.25\,nm $V$ amplitude. 
Gray and black show the network and inter-network, respectively.}
\end{figure}

\subsection{Correlations between photospheric and chromospheric quantities}

\subsubsection{Calcium core emission vs. magnetic flux}

The upper panel of Fig.~8 shows the relation between the photospheric magnetic flux and the 
chromospheric emission for the network. 
For lower flux values ($<$\,100 Mx\,cm$^{-2}$\,), there is a clear increase of emission with flux. 
However, for higher 
magnetic flux densities, the H-index increases slowly. 
To reproduce the observed relation, we utilize a power law fit to the data,
\begin{equation}
H\,=\,a\,\Phi^{b}\,+\,c,
\end{equation}
where $\Phi$ is the absolute magnetic flux density, $H$ is the H-index, 
$a$ is a constant coefficient, 
$b$ is the power index, and $c$ is the non--magnetic contribution (see section 5.3.2). 
We use a variable lower threshold of 0,\,3,\,..,\,20\,Mx\,cm$^{-2}$\, (cf. Table 3) 
for the fit and neglect all 
data points with fluxes below this level. 
We find that the power exponent, $b$, depends strongly on the threshold. 
The higher the threshold, the better the fit curve resembles a straight line, as first reported 
by  \cite{skumanich_etal_75}. 
If we keep all the points, including inter-network, we obtain  a value for the power exponent $b$ of $\approx$\,0.2.

\begin{figure}
\centering{\resizebox{8cm}{!}{\includegraphics{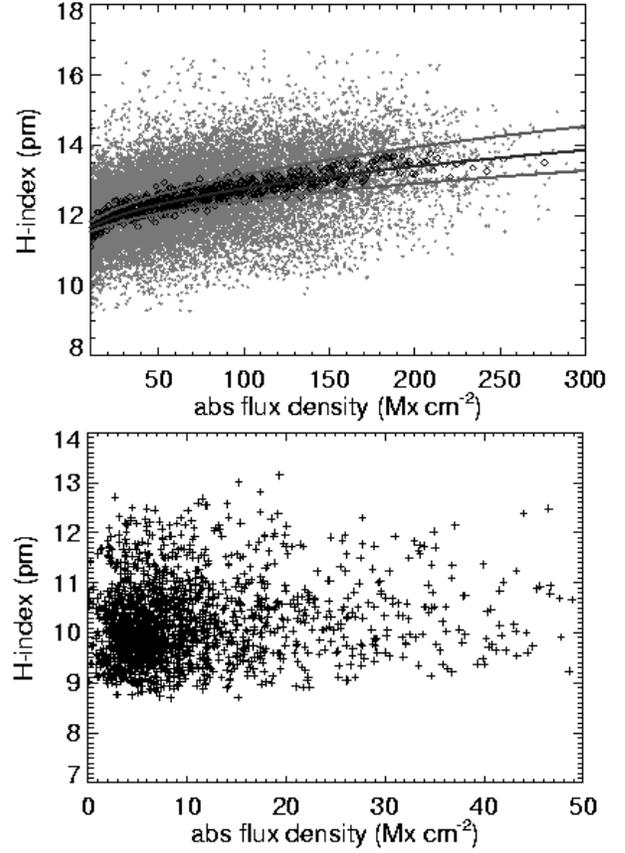}} }
\caption[]{{\em Upper panel}: Correlation between the H-index  and the absolute 
magnetic flux density. Gray is the original data and black is the binned data: each point is average of 25 points. 
The middle curve shows a fit of a power law to the original data (Eq. 1), the other two curves give the 
1$-\sigma$ error range of the fit. {\em Lower panel}:There is no correlation between the H-index and the 
magnetic flux density in inter-network.}
\end{figure}

\begin{figure*}
\resizebox{\hsize}{!}{
\includegraphics{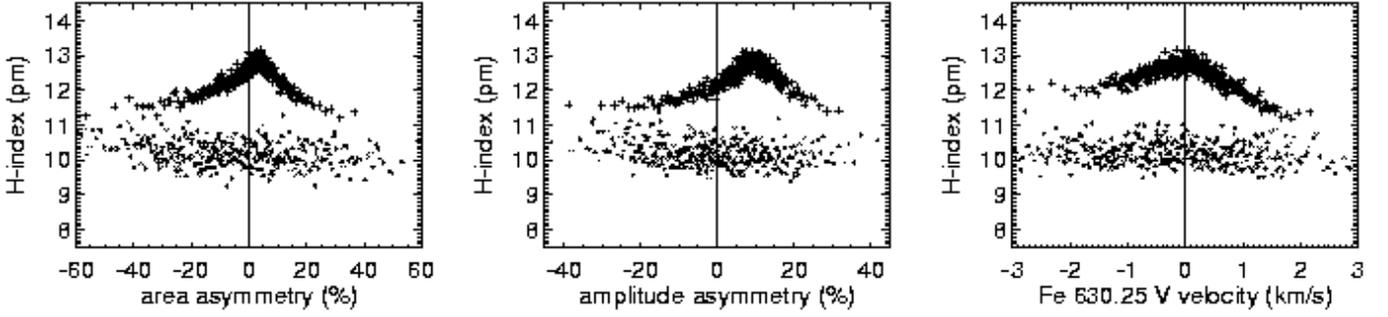}}
\caption[]{Correlation between the H-index and amplitude/area asymmetry and $V$ velocity. 
Plusses and squares show network and inter-network, respectively. The binning method is similar to Fig.~8}
\end{figure*}

The calcium core emission in the inter-network does not correlate with the magnetic flux, 
at least with the concentrated magnetic field within the Zeeman sensitivity (Fig.~8, lower panel). 
The value of the offset, $c$, of the fit to the network (Table 3) for low magnetic flux densities 
is around 10\,pm, which is consistent with the average H-index value of all the inter-network profiles
(10\,pm for the peak sample). The value of the $c$ parameter changes for different thresholds 
of the flux density (Table 3). Implications of this finding 
for the basal flux are discussed in Sect. 7.

\subsubsection{Calcium core emission vs. Stokes--$V$ velocity and area asymmetry}

There are  different behaviors for the network and inter-network  calcium core emission  
with respect to $V$ velocity and asymmetries (Fig.~9). 
The core emission peaks at small positive amplitude/area symmetry in the network.  
This is caused by strong $V$ profiles that show small asymmetries (Fig.~7, right panels).
The highest H-index in the network corresponds to almost zero $V$ velocity.  
In its diagram (Fig.~9), the slope of the left branch (blueshift) is smaller than the right one. 
Moreover for the strong upflow 
or downflow in the magnetic atmosphere in the network, the H-index decreases 
significantly. The H-index in the inter-network does not 
depend on any parameter of the Stokes--$V$ profile (Fig.~9).

Considering the fact that the amplitude and area asymmetries strongly 
correlate with each other (top left panel, Fig.~7), 
a natural consequence is that the left (right) branch of the scatter plot of the 
H-index vs. amplitude asymmetry corresponds to the left (right) branch of the scatter plot 
of the H-index vs. area asymmetry (Fig.~9).  
However, if we compare only negative or positive branches of the scatter plots of the H-index vs. $V$ 
velocity and asymmetries in the network, 
we realize that the left branch in the $V$ velocity \textit{does not} correspond to the 
similar branch in the asymmetry plots. 
The left diagrams in Fig.~10 consider only the profiles with a 
negative $V$ velocity, while in the right panels profiles with a negative area asymmetry are shown. 
A positive $V$ velocity may thus correspond to a positive or negative area asymmetry. 
There is no relation between the amplitude/area 
asymmetries and the $V$ signal in the inter-network (right panels, Fig.~7).

\begin{table}
\begin{center}
\caption{The parameters of the fit to Eq. 1 to the network data. 
The first column is the threshold for the magnetic flux density (Mx\,cm$^{-2}$\,). 
By increasing the threshold, we avoid the inter-network intrusions. 
For a comparison to the peak inter-network flux density, see Appendix A. }
\begin{tabular}{c c c c}\hline 
cut  &  $a$\,(pm)  & $b$ & $c$\,(pm) \\ \hline 
0   & 0.73$\pm$\,0.05 & 0.28$\pm$\,0.01 & 10.1$\pm$\,0.1\\\hline
3   & 0.71$\pm$\,0.05 & 0.29$\pm$\,0.01 & 10.1$\pm$\,0.1\\\hline
5   & 0.61$\pm$\,0.04 & 0.31$\pm$\,0.01 & 10.3$\pm$\,0.1\\\hline
10  & 0.22$\pm$\,0.02 & 0.45$\pm$\,0.02 & 11.0$\pm$\,0.1\\\hline
20  & 0.14$\pm$\,0.06 & 0.51$\pm$\,0.06 & 11.3$\pm$\,0.2\\\hline
\end{tabular} 
\end{center}
\end{table}

\section{The magnetically and non-magnetically heated components}

The H-index includes the H$_1$\,(the minima outside the emission peaks), H$_2$, and H$_3$ spectral regions. 
So its formation height extends from the higher photosphere to the middle chromosphere. 

The \ion{Ca}{ii}\,H\,\&\,K lines are one of the main sources of the chromospheric radiative loss.  
We use the H-index as a proxy for the  emission in the low/mid chromosphere, and 
assume a linear relation between the H-index and the chromospheric radiative loss. 
To derive the contribution of magnetic fields to the chromospheric emission, we
added the H-index of all points with magnetic flux above a given flux threshold. 
The \textit{fractional H-index}, $\eta$, is then defined by normalizing this quantity to the total 
H-index of all points in the field of view\footnote{We use all data points in this part 
and do not use the mask defined in Sect. 3.}:  
$$\eta \,(\Phi > \Phi_0) = \frac{\sum H \,\,\,\,\,(\Phi > \Phi_0)}{\sum H \,\,\,\,\,\,\ (\textrm{all maps})}.$$
Table 4 lists the fractional H-index in different magnetic flux thresholds. 
Some 20\,\% of the total H-index  is provided by strong flux 
concentrations ($\Phi\,\ge$\, 50. Mx\,cm$^{-2}$) while 
the remaining 80\,\% of that is produced by the weak magnetic field and/or field free regions.

\begin{figure}
\centering{\resizebox{8cm}{!}{\includegraphics{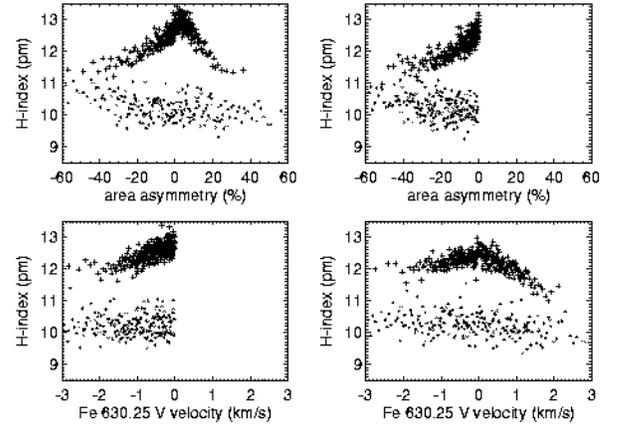} }}
\caption[]{In the left column diagrams, 
only the points with a negative $V$ velocity are plotted. 
In the right panels, only points with a negative area asymmetry are plotted. 
Pluses and squares show network and inter-network, respectively.}
\end{figure}
\begin{table}
\begin{center}
\caption{Fractional distribution of the total H-index, $\eta$, for different magnetic flux thresholds.}
\begin{tabular}{c c c}
$\Phi_0$\,(Mx\,cm$^{-2}$) & \# & $\eta$ (\%) \\ \hline 
4.   & 38\,905  & 63.7  \\\hline 
10.  & 26\,071  & 44.5  \\\hline 
50.  & 11\,190  & 20.4  \\\hline 
100. & 4\,933   & 9.2 \\\hline 
150. & 1\,709   & 3.3 \\ \hline 
200. & 329      & 0.6  \\\hline 
\end{tabular} 
\end{center}
\end{table}

The average H-index is 12.0 and 9.8\,pm in the network and inter-network, respectively. 
However, it contains some contributions from the photosphere (outside emission peaks) and a cool 
chromosphere (without temperature rise). 
Therefore, we decompose the observed calcium profiles in a 
\textit{heated} and a \textit{non-heated} component,  
\begin{equation}
\textrm{H}_{\textrm{i}} = \textrm{H}_{\textrm{co}} + \textrm{H}_{\textrm{non}} + \textrm{H}_{\textrm{str}}
\end{equation} 
where $\textrm{H}_{\textrm{i}}$ is the average H-index of the inter-network,  
$\textrm{H}_{\textrm{co}}$ is the non-heated component, $\textrm{H}_{\textrm{non}}$ is the 
non-magnetically heated component, and $\textrm{H}_{\textrm{str}}$ is the straylight contribution. 
The non-heated component of the profile originates from the line--wing emission (photosphere) 
and a cool chromosphere, $\textrm{H}_{\textrm{co}}$. 

We use the \cite{hol_mul_74} model atmosphere as a proxy 
for those parts of the solar atmosphere that are not affected by 
any heating process. 
This model is very similar to a theoretical radiative equilibrium 
model, but it was not constructed to strictly satisfy this condition.
We prefer this semi-empirical model over a theoretical 
radiative equilibrium model because it results in slightly better 
reproduction of the \ion{Ca}{ii}\,H line wings in a standard disk-center 
atlas.
We note, however, that 3D (M)HD simulations of the solar atmosphere 
commonly produce regions in the upper photosphere and chromosphere 
with temperatures significantly lower than in the Holweger-M\"uller model.
These structures, however, are generally so small that their evolution 
is strongly influenced by their surroundings, so that they cannot serve 
as a model for those parts of the solar atmosphere that are not influenced 
by any heating processes. That means that they are not suitable for determining 
the absolute minimum value of the H-index.

We perform a consistent NLTE radiative transfer computation in plane-parallel 
geometry, including effects of partial frequency redistribution, in order to 
estimate the non-heated component of the H-index. 
We added an isothermal hydrostatic extension up to a height of 2000\,km 
to the model in order to be able to compute the core parts of the \ion{Ca}{ii}\,H\,\&\,K lines. 
This may seem an arbitrary choice, but considering that the computed 
intensity is essentially zero at the line core for models with such low 
temperatures in the upper atmosphere, its influence on the H-index 
is very small. This model atmosphere and the obtained profile  are 
similar to the COOLC model of \cite{ayres_etal_86} 
and the corresponding profile. 
 We find a value of $\textrm{H}_{\textrm{co}}\,= 5$\,pm at disk center. 
At a heliocentric angle of 53\degr, 
the appropriate value is $\textrm{H}_{\textrm{co}}\,= 5.9$\,pm.

We find an upper limit of $\sim\,12\,\%$ for the straylight contamination of the 
observed profiles (cf. Appendix B).
Therefore, we decompose the average H-index in the inter-network ($\textrm{H}_{\textrm{i}}\,=\,9.8$\,pm) to obtain 
the non-magnetically  heated component, $\textrm{H}_{\textrm{non}}$:

\begin{equation}
\textrm{H}_{\textrm{non}} = \textrm{H}_{\textrm{i}} - \textrm{H}_{\textrm{co}} - \textrm{H}_{\textrm{str}}
\end{equation} 
\begin{footnotesize}\begin{eqnarray*}
\textrm{H}_{\textrm{non}}  = 9.8\,(\pm\,1.0) - 5.9\,\left(^{+0.6} _{-0.0}\right) - 1.1\,(\pm\,0.1) = 2.8\,\left(^{+1.0} _{-1.1}\right).
\end{eqnarray*}
\end{footnotesize}
We find an average non-magnetic heating of $\textrm{H}_{\textrm{non}}\sim\,2.8\,\pm\,1.1\,$pm. 
To decompose the average H-index of the network, we also consider the magnetically heated component, 
$\textrm{H}_{\textrm{mag}}$, and use the same method:
\begin{equation}
\textrm{H}_{\textrm{mag}} = \textrm{H}_{\textrm{n}} - \textrm{H}_{\textrm{co}} - 
\textrm{H}_{\textrm{non}} - \textrm{H}_{\textrm{str}}.
\end{equation} 
where the magnetically heated contribution is the first term in Eq. 1. 
Assuming a similar non-magnetically heated contribution, $\textrm{H}_{\textrm{non}}$, for the network as in the 
inter-network~\cite[which is supported by observations, e.g.,][]{schrijver_XI_87, schrijver_95}, 
we find an average magnetically heated component of about $\textrm{H}_{\textrm{mag}}\sim\,1.9\,\pm\,1.4$\,pm, 
which is consistent with the difference between 
the mean H-index in the network and inter-network. 
Thus, the non-magnetically heated component has a larger contribution in the observed H-index 
than magnetically heated one, both in the network and inter-network. 
Our error estimates in Appendix B show that the uncertainties of the measurements do not 
affect this result.

\section{Discussion}
\subsection{Thermal coupling}
All intensity parameters, from the calcium core to the \ion{Fe}{i}\,630\,nm continuum, 
show different distributions in the network and inter-network: the network patches are 
brighter not only at all observed wavelengths of the calcium profile, but also in the 630\,nm 
continuum (Figs.~7\,and\,4). 
As noted by \cite{cram_dame_83}, this implies that the temperature fluctuations are spatially 
coherent between the photosphere and the lower chromosphere.   
This coherent intensity increase is presumably due to the fact that  
the optical depth scale is shifted downwards in the evacuated magnetic part of the atmosphere. 

The correlation  between the inner wing intensity and H-index is strong, 
both in the network and inter-network (Table 2). However, the correlation between 
the outer wing intensity and the H-index is significant only in the network (Fig.~6, lower left panel). 
This indicates that the height range of the thermal coupling between the photosphere 
and low/mid chromosphere increases in presence of a magnetic field. 
There are also suggestions that this coupling extends to the upper chromosphere~\citep{rauscher_marcy_06}.

\subsection{H-index vs. magnetic flux}
The amount of energy deposited  in the network depends on the 
magnetic flux density. To reproduce the dependence of calcium 
core emission on the magnetic flux with a power law, we find that 
the exponent derived strongly depends on the inclusion (or exclusion) 
of the weakest fluxes. We obtain a power 
exponent, b, of about 0.3,  which is smaller than the value given 
by~\cite{schrijver89} and~\cite{harvey_white_99}. We ascribe the 
difference to the higher quality of our data set: better spatial 
resolution, a lower detection limit for magnetic signals, an accurate 
estimate of the magnetic flux due to the inversion of vector polarimetric 
data instead of a magnetogram, and unlike these authors, we do not 
consider an assumption for the background component prior to the fit. 
Instead, we keep it as a free 
parameter. In this way, we find a background component close to the mean H-index of the inter-network for 
the weakest flux concentrations.

The rate of increase of the chromospheric emission vs. magnetic flux 
reduces for strong magnetic flux densities. This implies that for large 
flux concentrations, either the filling factor or the magnetic field strength saturates. 
There are indications that it is the filling factor:  
the available space is completely filled by expanding flux tubes~\citep{hammer87, saar_96, schrijver96}. 
In contrast,  a lower limit of chromospheric emission also exists. 
In the inter-network, no relation between emission and the photospheric fields is found (Figs.~8\,and\,9). 
The average value of the H-index in the inter-network of around 10\,pm corresponds to the offset, $c$, Eq. 1. 
This reflects a constant contribution  to the H-index which is present even 
without photospheric magnetic flux, in agreement with 
\cite{schrijver_XI_87, schrijver_95} who argued that 
the basal flux does not depend on the magnetic activity. 
The basal flux contains two components:  the non-heated (cf. Sect. 6)  and the non-magnetically heated contributions. 
The non-heated contribution depends on the temperature stratification, 
so that we speculate that the non-magnetically heated component  
has an inverse dependence on the temperature stratification.

Figure 8 (upper panel) shows that there is a variable lower limit for the H-index versus the 
magnetic flux density. 
In contrast, the upper limit is less clearly defined, and would be in
agreement with a constant maximum value independent of the amount of magnetic flux.
There are also similar behaviors for the upper and lower limits in the scatter plots of 
the H$_3$, H$_{2\textrm{v}}$, and H$_{2\textrm{r}}$ versus the magnetic flux density. 
This is similar to upper and lower limits of the basal flux of stars vs. the color (B-V) where 
the lower boundary changes but the upper one is almost constant~\cite[][their Fig.~2]{fawzy_etal_02b}. 
The situation for the inter-network is different: both the upper and lower limits  are 
independent of the magnetic flux density. 

\subsection{Decomposing the \ion{Ca}{ii}\,H profile}
We interpret the solar \ion{Ca}{ii}\,H line profile as the superposition of a cool chromospheric profile, 
the heated component and straylight components. 
\cite{oranje_83} first applied this method to reproduce different observed Ca\,II\,K profiles from two 
``basic'' profiles. 
Later, \cite{sol_stein91} used the same idea to decompose the observed profiles by a 
combination of two theoretically calculated profiles. 
Our one-dimensional NLTE radiative transfer calculations reveal that 
the H-index of a cool chromosphere is about 5.9\,pm.  
Since the chromospheric emission of the Sun as a star is slightly above the minimum among the 
Sun--like stars~\citep{schrij_zwaan}, 
the average quiet Sun  profile contains a heated component. 
Subtraction of the cool chromospheric component from the measured H-index 
(along with straylight considerations) leads to an estimation of 
the \textit{pure} non-magnetically and magnetically heated components in the measured H-index. 
The non-magnetically heated component is  almost 50\,\% larger than the magnetically heated component.  
The presence of a significant non-heated contribution in  the H-index indicates 
that not all of the chromospheric emission emerges from a \textit{hot} chromosphere, 
in contrast to findings from, e.g., \cite{kalkofen_etal_99}. 
The minimum  heated component in a Ca profile provides evidence how cool the chromosphere may 
be~\citep[e.g.,][]{wedemeyer_etal_05}. 
Moreover, it favors theories in which the non-magnetic heating plays  
the dominant role in the chromospheric heating~\citep{fawzy_etal_02a}. 

The fact that the non-magnetic chromospheric heating contributes significantly to 
the chromospheric energy balance is also found in the relative contributions of mainly 
field--free inter-network and network areas to the total emission in the field of view. 
Magnetic flux densities above 50\,Mx\,cm$^{-2}$ add 
only 20\,\% to the total emission, which is also seen in the ratio of the mean H-index 
in the network and inter-network, 12.0\,/\,9.8 $\approx$\,1.2 (Table 4) .

\subsection{Uncertainties}
The main uncertainty in the  non-magnetically and magnetically heated components is the presence  
of some heating in our cool chromosphere profile.  
The \cite{hol_mul_74} model atmosphere is close to an atmosphere in radiative equilibrium without 
non-magnetic or magnetic heating, so no heated component (cf. Sect. 6) is expected 
to be found on its Ca profile. 
Since there is no general agreement how cool the chromosphere may 
be~\citep[e.g.,][]{kalkofen_etal_99,ayres_02}, 
the best answer would be to use the observed calcium profile with the lowest H-index. 
However, \cite{sol_stein91} concluded that it is not possible to observe a low activity 
profile as presented by the COOLC model of \cite{ayres_etal_86}, which is similar to the one employed by 
us on base of the \cite{hol_mul_74} model atmosphere. Thus, our cool profile can 
well serve as lower limit of the non-heated contribution to the H-index.  
Note that a larger value for $\textrm{H}_{\textrm{co}}$  would increase 
the contribution of the magnetically heated component, leading to a larger fractional contribution  
for the magnetically heated to the total heated component.

There may be some mixed polarity fields below our polarimetric detection limit which may 
influence the ratio of the magnetically heated to the non-magnetically heated component. 
  
\subsection{H-index vs. $V$ asymmetries}
From the relations of the chromospheric emission to quantities of the photospheric 
magnetic field in the network, we find that the chromospheric emission peaks at 
small positive values of the Stokes $V$ asymmetries and at zero Stokes $V$ velocity 
(Fig.~5). It reflects the histograms of the respective photospheric field quantities 
which show similar distributions even at disk center~\citep[e.g.,][]{sigwarth99}. 
However, in combination with the dependence of the emission on the flux density it is 
inferred that the stronger flux concentrations mainly show small material flows  
along with non-zero asymmetries. 

The magnetically heated component is related to Stokes $V$ profiles with non-zero area and 
amplitude asymmetries (Figs.~5\,and\,9). 
A possible explanation for this finding would be the absorption of upward propagating 
 acoustic waves, generated by the turbulent convection, by the inclined fields of expanding 
flux tubes. This would imply that the energy is deposited  at the outer boundary 
or in the canopy of flux concentrations rather than in the central, more vertical, part. 
It is mainly because the non-magnetic cutoff frequency is lowered at the boundary of the flux tubes, 
where the field lines are inclined. 
This was first predicted by~\cite{suematsu_90} and recently achieved some observational 
support~\citep{hansteen_etal_06, jefferies_etal_06}. 
Figure 9 indicates that the maximum observed H-index has non-zero $V$ asymmetry. 
These asymmetric $V$ profiles are consistent with the case when the line of sight passes through the canopy of 
a magnetic element or through a flux tube axis~\citep[positive and negative asymmetries respectively,][]{steiner99}. 
Therefore our finding supports~\cite{suematsu_90}.

\subsection{Inter-network field strength distribution}
As a byproduct of our study on the relation between photospheric fields and chromospheric emission, 
we obtained distributions of field strength for network and inter-network regions. For the inter-network fields, 
for the first time an inversion of visible spectral lines in the weak field limit led to the 
same distribution as results from the more sensitive infrared lines \citep[][]{collados_01}. 
This is discussed in Appendix A. Thus, we find that the inter-network magnetic field is dominated by 
field strengths of the weak field regime which ranges up to some 600\,G.
This rules out the argument that the inter-network is dominated by a 
magnetic field with a strength of more than a kilo-Gauss~\citep[e.g.,][]{almeida_cerdena_kneer03}. 

\section{Conclusions}

There is no correlation between the H-index  and magnetic field parameters in the inter-network. 
Therefore, we conclude that the magnetic field has a negligible role in 
the chromospheric heating in the inter-network. 
On the other hand, the H-index is a power law function of the magnetic flux density 
in the network with a power index of 0.3. 
The average H-index observed in the network and inter-network are $\sim$\,10 and 12\,pm, respectively. 
We find a non--magnetic component in the network (based on Eq. 1) of about  10\,pm which 
shows the consistency of our analysis: the non--magnetic part of the 
network H-index 
is equal to the H-index of the inter-network. 

A NLTE radiative transfer calculation, using the Holweger--M\"uller model atmosphere, 
indicates that the non-heated component of the H-index, emerging from a cool 
chromosphere, is about  5.9\,pm. 
Comparison of this non-heated component and the average H-index in the inter-network 
has two implications: a) some of the observed chromospheric emission does not originate from a hot chromosphere, 
and b) the non-magnetically heated component is  about 50\,\% larger than the magnetically heated component. 
From this, we conclude that the non-magnetically heated component has a larger contribution 
in the chromospheric radiative loss than 
the 
magnetically  heated component, both in the network and inter-network. 

In our statistical ensemble, 
spatial positions with strong magnetic field ($\Phi\ge$\,50 Mx\,cm$^{-2}$) contribute about 20\,\% 
of the total H-index. Correlations and histograms of the different intensity bands in the \ion{Ca}{ii}\,H spectrum 
indicate that above a magnetic threshold, photosphere and low/mid chromosphere are thermally 
coupled. Moreover, our findings are consistent with the idea that the energy transfer in a flux tube 
has a skin effect: the energy transfer is more efficient in the flux tube boundary (canopy) 
than at its center (axis). 

For the first time, we find a magnetic field distribution in the inter-network using visible lines 
which is similar to results inferred from infrared lines with larger Zeeman splitting. 
The distribution function increases with decreasing field strength. The peak at 200\,G is due to the detection
limit of the polarization signal. The average 
and distribution peak of the absolute flux density (of magnetic profiles) 
in the inter-network are $\sim$\,11 and 4 Mx\,cm$^{-2}$, respectively. 
We conclude that the combination of high spatial resolution and polarimetric accuracy is sufficient 
to reconcile the different results on field strength found from infrared and visible lines. 

\begin{acknowledgements}
The principal investigator (PI) of the ITP observing campaign was P. S\"utterlin, Utrecht, The Netherlands. 
We wish to thank Reiner Hammer, Oskar Steiner, Thomas Kentischer, Wolfgang Rammacher, and Hector Socas-Navarro
for useful discussions.
The POLIS instrument has been a joint development of the High  
Altitude Observatory (Boulder, USA) and the Kiepenheuer-Institut. 
Part of this work was supported by the Deutsche Forschungsgemeinschaft (SCHM 1168/8-1).
\end{acknowledgements} 

\bibliography{rezabib}        
\appendix

\section{The magnetic field distribution}
Traditional magnetic field strength distributions for the quiet Sun (inter-network) based on visible lines 
have a broad peak centered around 1\,kG~\citep{almeida_cerdena_kneer03,cerdena_almeida_kneer06}. 
On the other hand, observations in the infrared lines show a clear peak at 
lower field strengths and an almost exponentially 
decreasing probability for higher fields strengths~\citep{collados_01,luis_collados_03,khomenko_etal_03}. 
This discrepancy between the retrieved parameters from the visible and infrared lines 
has not been solved so far~\citep{steiner_02,martinez_etal_06}.

We achieved a magnetic field distribution for the inter-network 
using the \ion{Fe}{i}\,630\,nm pair that is similar to the infrared studies (Fig.~5.b). 
We elaborate on the effects of spatial resolution and signal-to-noise ratio on this finding.  
We used the Kiepenheuer Adaptive Optics System (KAOS) to improve spatial 
resolution and image stability~\citep{luhe_etal_03}. 
It  provides  stable sharp images in a certain field of view, and therefore   
allows us to increase the exposure time to have better signal-to-noise ratio. 

Figure A.1 is a close-up view of the inter-network magnetic flux distribution. 
The lowest flux detected is $\sim$\,0.05\,Mx\,cm$^{-2}$\,. { The distribution of the 
flux decreases sharply for fluxes below $\Phi\,\le\,4$\,Mx\,cm$^{-2}$, which is due 
to the detection limit of the polarization signal. The small fluxes still detected 
give us, however, confidence that the magnetic field strength distribution of Fig.~5.b 
is reliable as a consequence of the combination of  high spatial resolution and good signal-to-noise ratio.}

\begin{figure}
\resizebox{\hsize}{!}{\includegraphics{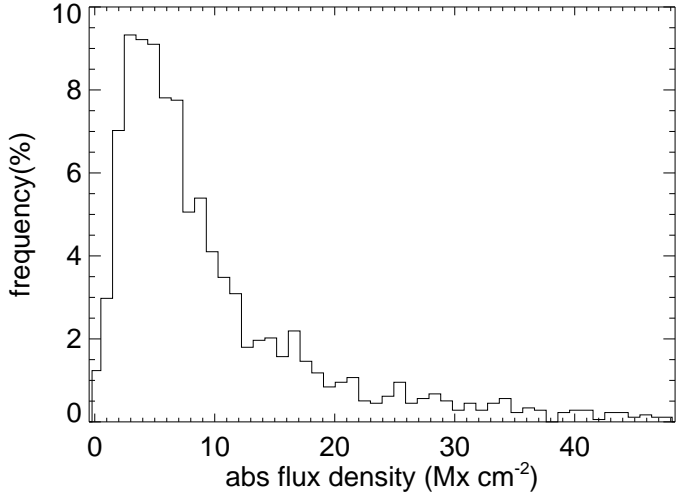}}
\caption[]{ Close up view of the flux density distribution for the inter-network. 
At a value of $\sim$\,4.0\,Mx\,cm$^{-2}$\,, the histogram shows a peak.}
\end{figure}

To investigate the influence of noise on the inversion results, 
we compare inversion results of the original datasets with and without adding noise. 
In order to create ``noisy'' data sets, the rms value of the noise in the original data 
was increased to $\sigma$\,=\,15.0\,$\times 10^{-4}$\,$I_c$,  
two times larger than the original value. 
The spatial sampling was  identical  to the original data. 
We emphasize that inversion was done using the same assumptions and initial model atmosphere.  
Figure A.2 compares two sets of full Stokes profiles (black) 
and the inversion fit (red). The upper four panels are the original profiles and the lower four 
panels are the same profiles with noise. The fits and retrieved model atmospheres differ significantly. 
Especially important is the fact that the {\em  noisy} inversion leads to a magnetic field of 1.5\,kG.

\begin{figure}
\resizebox{\hsize}{!}{\includegraphics{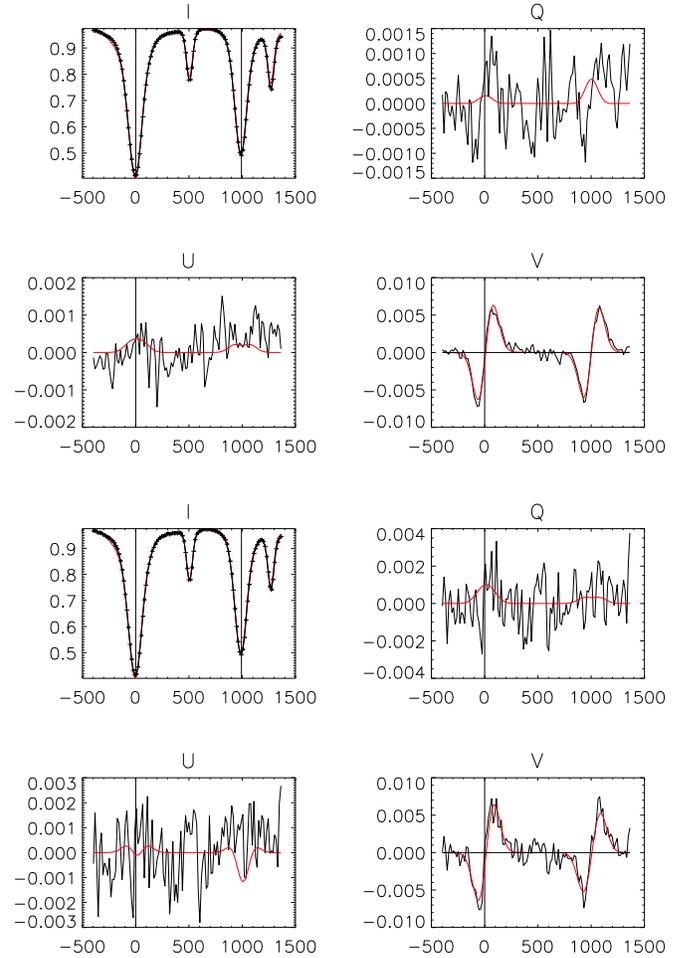}}
\caption[]{The upper four plots show an example of the original Stokes profiles and the inversion fits. 
The lower four figures show the same profile with a rms noise value two times larger and corresponding fits. 
The magnetic field strength, magnetic filling factor, and macro turbulence velocity for the 
original and noisy profiles are (842, 1543) G, (7.6, 5.3)\%, and (2.95, 2.77) km\,s$^{-1}$ respectively. 
The x-axis shows the spectral coordinate in m\AA.
}
\end{figure}

To compare the field strength obtained from the original and noisy datasets, their 
histograms are shown in Fig.~A.3. The distribution of the original field strength 
peaks around $\sim$\,200\,G (black histogram), while for the noisy data, it has a clear shift 
toward higher values with a peak at 0.8\,kG (red histogram). 
{ This emphasizes the} role of the noise in the existing discrepancy between visible and 
infrared measurements~\citep{luis_collados_03}. 

In brief, we find that higher spatial resolution of these observations along with low noise data 
is an important step toward resolving disagreements between the visible and infrared polarimetric measurements. 
The visible \ion{Fe}{i}\,630\,nm pair  shows higher Stokes amplitudes for small magnetic fields than 
the 
infrared lines of \ion{Fe}{i}\,1.56 $\mu$m. Hence, with an equal amount of noise in observational data, 
the visible lines provide a higher signal-to-noise ratio than infrared lines. 
Therefore, in contrast to ~\cite{martinez_etal_06}, we find the visible \ion{Fe}{i}\,630\,nm pair to be a proper 
tool to investigate the inter-network magnetic field. We  ascribe the present disagreements 
to the low spatial resolution of previous observations and different signal-to-noise ratios.

\begin{figure}
\resizebox{\hsize}{!}{\includegraphics{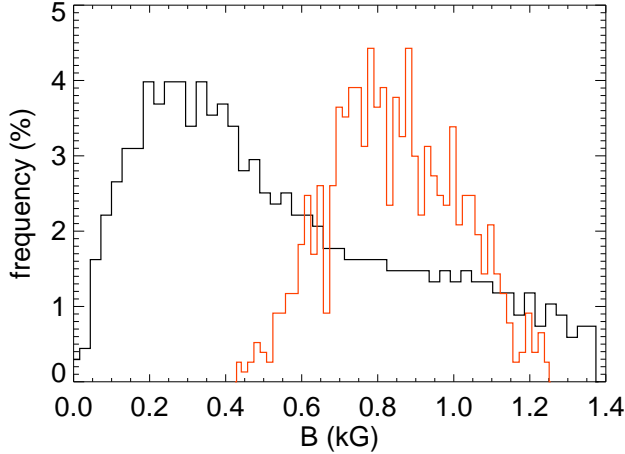}}
\caption[]{Histograms of the inter-network field strengths of one of the maps 
based on the inversion of original data (black) and noisy data (red).}
\end{figure}

\section{Calibration uncertainties in POLIS Ca channel}

The POlarimetric LIttrow Spectrograph (POLIS) was designed to 
facilitate co--temporal and co--spatial measurements of the vector magnetic field in the photosphere 
and the \ion{Ca}{ii}\,H intensity profile\citep{schmidt03,beck05b}. 
We try to elaborate  on the  uncertainties in 
 the intensity calibration, because it is critical for the estimates of 
the various contributions (straylight, non-magnetic heating, etc.) to the observed calcium profiles.
Observations are discussed in Sect. B.1. 
We study  the linearity of the CCD camera for different light levels in Sect. B.2. 
Then, we investigate instrumental and solar straylight contributions in Sect. B.3. 
In Sect. B.4, we perform standard error propagation to quantify uncertainties 
in our final calibrated spectra.

\begin{figure}
\resizebox{\hsize}{!}{\includegraphics{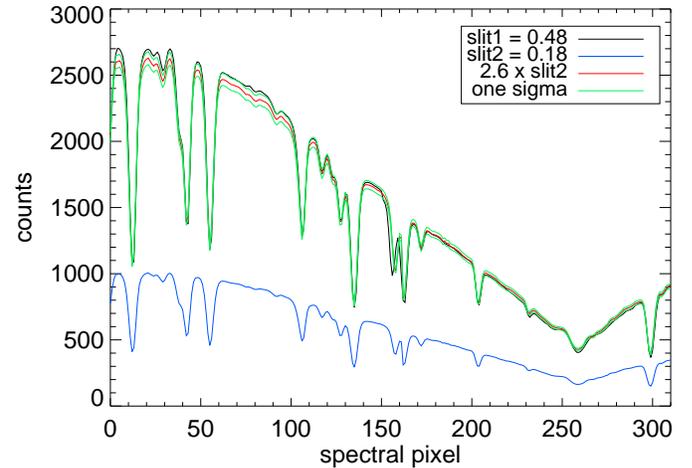}}
\caption[]{Comparison between two different slit widths and the same modulation speed. 
The black and blue profiles 
correspond to slit widths of 0.48 and 0.18 arcsec, respectively. The red profile is 
the blue profile multiplied by a constant factor of 2.6 for all pixels. 
The green profiles show the  1-$\sigma$ RMS value ($\sim$\,1\%) from the mean curve (red).} 
\end{figure}

\begin{figure}
\resizebox{\hsize}{!}
{\includegraphics*{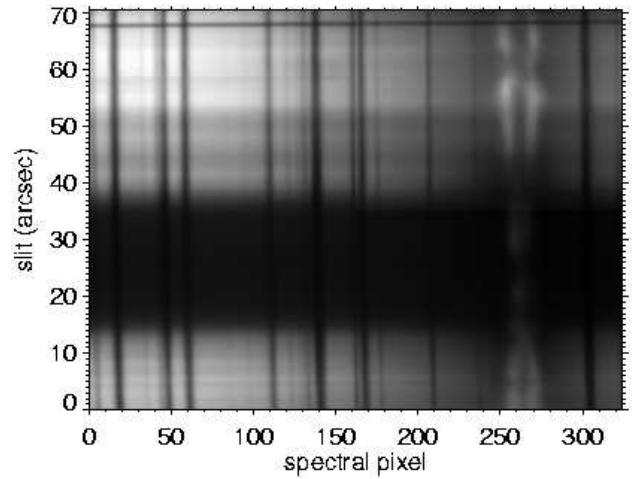}}
\caption[]{Dark subtracted spectrum of the umbra in a sunspot.}
\end{figure}

\begin{figure*}
\resizebox{\hsize}{!}
{\includegraphics*{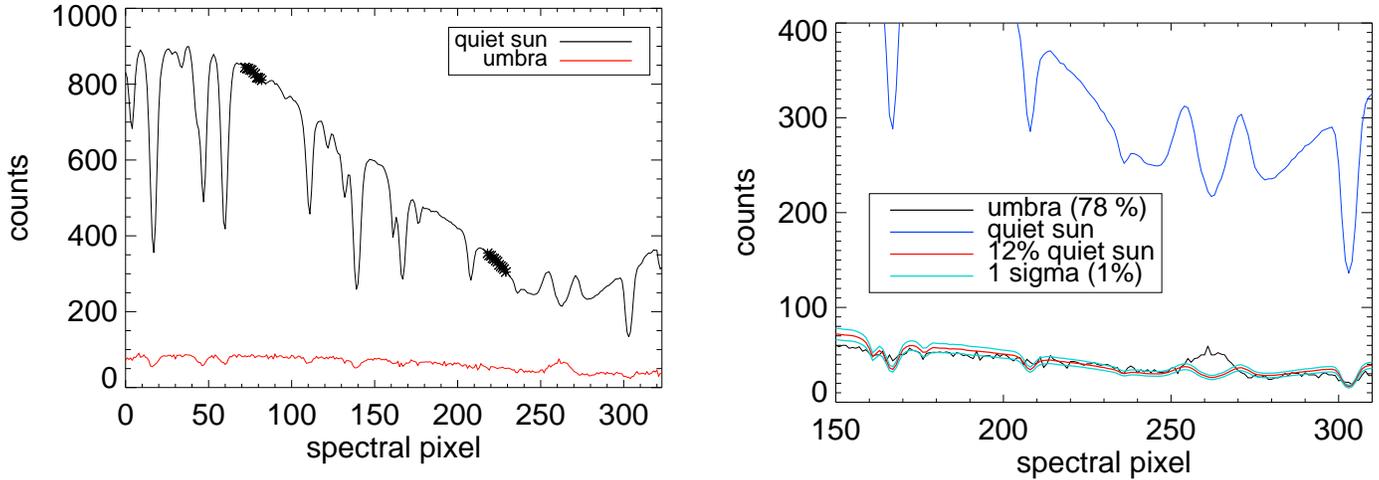}} 
\caption[]{{\emph Left:}  a single \ion{Ca}{ii}\,H spectrum of the umbra and quiet sun profiles in a sunspot. 
Asterisks show spectral bands that were used to estimate the true signal (see Sect. B.3.2). 
{\emph Right:} close up view of the quiet sun and umbral profiles near the calcium line core. 
The true signal (about 5\,\% in the wing close to the Ca core according 
to \cite{umbral_atlas_fts} was subtracted from the umbral profile. 
The red curve shows 12\,\% of the quiet sun profile. 
The two cyan curves show an uncertainty interval of one sigma (1\,\%).}
\end{figure*}

\subsection{Observations}
In order to study linearity of  the POLIS  CCD camera of the calcium channel 
at different light levels, we observed 
a set of calcium spectra and flat field  
data with different accumulations and slit widths at  disk center. 
A sunspot was also observed with a similar setup and an exposure time of 0.82\,s and six accumulations. 
Besides this, 
we recorded  dark current data using  either the AO or POLIS field stop. We use these two data sets to 
estimate the solar and instrumental straylight, respectively. 

All spectra and profiles in this Appendix are dark subtracted raw data. 
No other calibration process like flat fielding was applied. 
In order to show intrinsic noise in the data, we did not  spatially average darks or spectra.

\subsection{Linearity} 
In order to check the linearity of the calcium  camera from profiles 
by, e.g., comparing the ratio of the core to wing for different exposures, 
we recorded spectra with different slit widths and identical modulation 
speed \citep{ccd_char_94}.  
Figure B.1 shows a comparison between the two flat field profiles taken on disk center 
with identical illumination, but different slit widths. 
The black and blue profiles correspond to  a 
slit width of 0.48 and 0.18 arcsec, respectively. 
If one multiplies the blue curve  with the ratio of the slit widths of 2.67, 
the resulting curve agrees well with the profile of the broad slit for all wavelengths 
including the line core. 
We calculated the ratio of the two profiles with different slit widths as function of wavelength. 
The rms variation of the ratio was 1\%. 
This can be used as an upper limit for non-linearity effects of the CCD camera.

\begin{figure}
\resizebox{\hsize}{!}
{\includegraphics{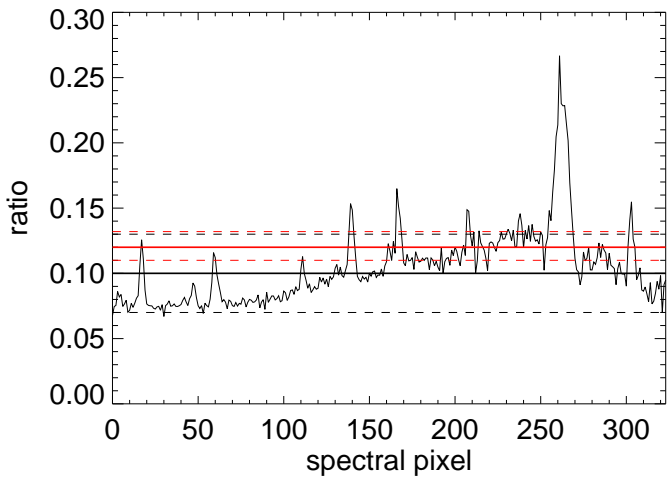}} 
\caption[]{Ratio of the umbral profile to a quiet sun profile. 
If one considers all the spectrum, the ratio will be 10\,$\pm$\,3\,\% while considering only parts 
of the wing close to the \ion{Ca}{ii}\,H core, the ratio will be 12\,$\pm$\,1\,\%. Note to the 
mismatch between the quiet sun and umbral profiles at spectral line positions due to different broadenings.}
\end{figure}

\subsection{Straylight} 

\subsubsection{False light}
We compare two sets of dark current  images  of  the POLIS Fe channel: 
one with  a field stop right after the focal plane,  
and  another one with the field stop in front of the camera.  In the second case, 
all light sources inside the observing room are fully blocked.  The difference 
between the two measurements  is about 0.9\,$\pm$\,0.5\,\%. This  gives an upper 
limit for  the instrumental straylight in the POLIS Ca channel as well, 
since most of the sources in the observing room  will not emit in blue as efficient as red.

\subsubsection{Atmospheric and telescopic straylight}
The calcium profile in the sunspot umbra is significantly weaker than a quiet sun profile. 
However, there is a well known single emission peak  in the calcium core~\citep{linskyavrett}. 
We  thus compare an inner wing band close to the calcium core, but excluding the 
core itself (asterisks in the left panel of Fig.~B.3), 
in an umbral profile and  a normal profile  from outside the spot to derive an 
upper limit for the solar straylight  contribution to observed profiles.

Figure B.2 shows a slit spectrum of the umbra. 
To keep the sample profiles as uniform as possible, we chose two identical positions  
along the slit (at about 30 arcsec, Fig.~B.2). 
The spectrum of the quiet sun and the 
minimum spectrum of the umbra are shown in Fig.~B.3 (left panel).

\begin{figure}
\resizebox{\hsize}{!}{\includegraphics{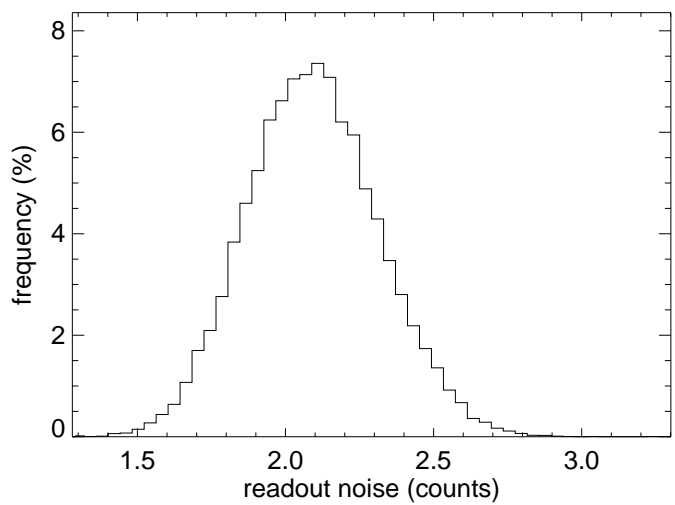}}
\caption[]{Distribution of the readout error in a set of 50 dark currents with 8 accumulations. }
\end{figure}

\cite{umbral_atlas_fts} presented a Fourier transform spectral atlas of the umbra. 
In this atlas, the wing close to the umbral core 
has an intensity of 5\,\% in units of the umbral continuum,  
while the wing close to the \ion{Fe}{i}, ]ion{Ti}{i}, and \ion{Ni}{i} lines at about 
396.5\,nm has an intensity of 38\,\%. In our umbral profile, 
this continuum has an intensity of about 80 counts (asterisks, Fig.~B.3, left panel), 
whereas close to the core the profile has about 45 counts. 
Assuming no straylight in the atlas profile, only 11 counts ($\sim$\,22\,\% of the observed valued) 
should be measured in the core. Thus, 78\% of the observed signal have to be due 
to straylight. 
To reproduce 78\% of the umbral profile, a straylight contribution of around 12\,\% is needed.
If we assume that there is no real calcium signal in the umbra (which is wrong) and 
take the  observed calcium profile of the umbra as pure scattered light, 
the ratio changes from 12\,\% to 15\,\%. 
Therefore, uncertainties about the scattered light in the atlas profile have minor importance.

A close-up view of the spectral region close to the calcium core is shown in Fig.~B.3 (right panel). 
For the reason mentioned above, 
in the right panel of Fig.~B.3, the umbral profile was labeled \textit{umbra (78\,\%)}. 
The red curve shows 12\,\% of the quiet sun profile. The two cyan curves also show 
an uncertainty interval of one sigma (1\,\%). Figure B.4 shows the 
ratio of the umbra to the quiet sun profile. If we calculate mean and RMS of the whole 
spectrum, it gives a value of 10\,$\pm$\,3\,\% while considering only on the wing band  
close to the calcium core, we obtain 12\,$\pm$\,1\,\%. We take this value as an upper limit 
for the total straylight in the POLIS calcium channel. 
In Sect. B.3.1, we concluded that the instrumental straylight is about 1\,\%. 
Hence, most of the obtained straylight originates from 
the scattered solar light in the telescope and the earth atmosphere.

\begin{figure}
\resizebox{\hsize}{!}{\includegraphics{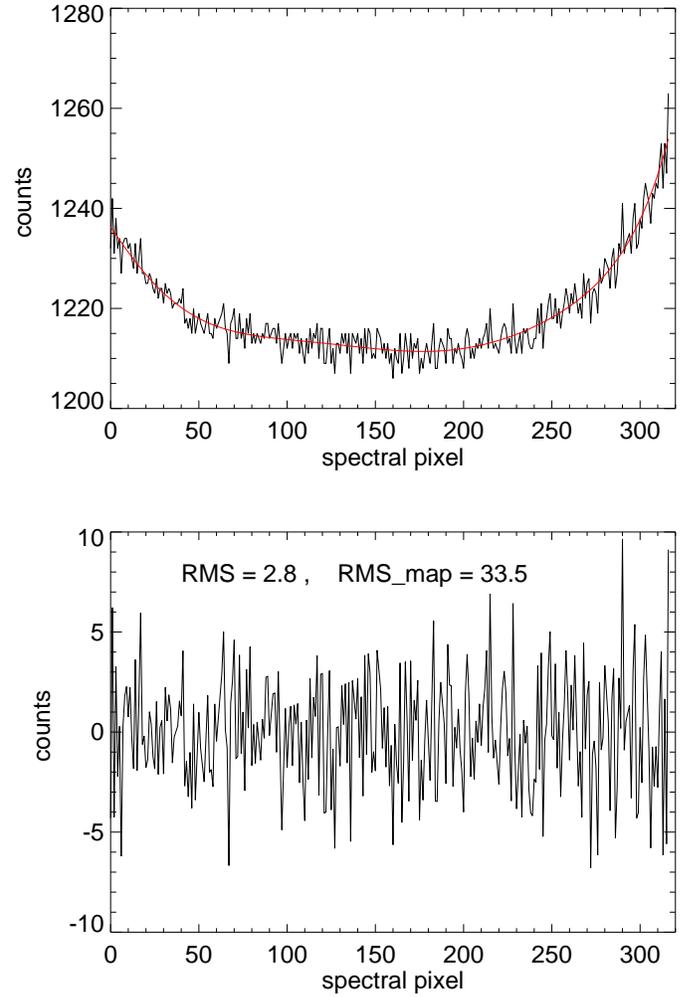}}
\caption[]{{\emph Upper panel:} a row in a dark frame with 6 accumulation. The red curve is a polynomial fit 
to subtract systematic shift from the random noise. 
{\emph Lower panel:} The residual signal after subtraction of the polynomial, which is the real noise. 
Note to the large difference between the rms of the map (excluding its borders) and the rms of the residuals.}
\end{figure}

\subsection{Noise}

There are three noise sources in a CCD camera~\citep{ccd_char_94}:
\begin{enumerate}
\item readout noise because of the pre-amplifier

\item dark current noise which is influenced by the chip temperature

\item photon noise due to the quantum nature of light which obeys Poisson statistics.
\end{enumerate}

The readout error of the camera was estimated by a set of 50 consecutive dark frames 
with similar exposure and accumulations. 
Here, we compare  the number of counts on each pixel  in different  images. 
We attribute the  standard deviation of the count value mainly to the readout 
error, and hence, obtain a readout error for each pixel separately.
The distribution of the  readout errors is shown in Fig.~B.5. 
The readout noise is about $\sigma_{\textrm{readout}}\,\sim\,2$ counts.

To investigate the thermal noise in the dark current, one has to 
remove small systematic offsets between CCD columns 
(probably due to thermal fluctuations of the AD system) before 
considering statistical measures, e.g., rms, of a dark profile (a CCD row). 
As it is shown in Fig.~B.6, for a dark frame with 6 accumulations, 
the resulting rms (after removing systematics) 
is far less than the standard deviation of the whole map including a large-scale variation. 
It provides an estimate for the thermal noise in the camera.  
So the rms of the dark frame is about $\sigma_{\textrm{thermal}}\,\sim\,3$ counts.

The total noise in the recorded data, $\sigma_{\textrm{noise}}$, consists of three components:
\begin{equation}
\sigma_{\textrm{noise}}=\sqrt{\sigma_{\textrm{photon}}^2+\sigma_{\textrm{thermal}}^2+\sigma_{\textrm{readout}}^2}
\end{equation}
where the photon noise originates from the Poisson fluctuations of the number of photons counted. 
Assuming that n$_{\textrm{photon}}$ is the number of detected photons, 
the corresponding rms, $\sigma_{\textrm{photon}}^2 = \textrm{n}_{\textrm{photon}}$. 
The calcium core is more influenced by the noise than the other points in the spectrum. 
Figure B.7 shows a dark subtracted calcium spectrum with 6 accumulations and a narrow slit (0.18 arcsec). 
The minimum number of counts in the core is 117, so $\sigma_{\textrm{photon}} \sim 11$. 
Therefore, the total noise in the core (Eq. B.1) will be $\sigma_{\textrm{noise}}\sim 11$. 
So the worst signal-to-noise ratio at the calcium core position will be better than 10. 
Note that the data described in the paper were obtained with a wide slit (0.48 arcsec)
while the spectrum shown in Fig.~B.7 was recorded with a narrow slit (0.18 arcsec),  
so we have better signal-to-noise ratio in the data.

\begin{figure}
\resizebox{\hsize}{!}{\includegraphics{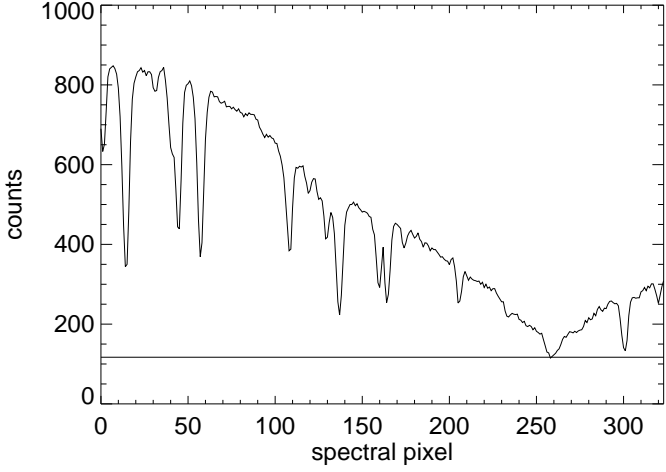}}
\caption[]{A sample calcium profile.  We use the calcium core minimum  counts to estimate the photon noise. 
In this case, the signal-to-noise ratio at the calcium core is more than 10.}
\end{figure}

\subsection{Error propagation} 

The relative intensity of the core to wing is defined as:

\begin{equation}
x = \frac{\textrm{I}_{\textrm{core}}}{\textrm{I}_{\textrm{wing}}} 
\end{equation} 

To estimate their RMS errors, we find the signal-to-noise ratio for the core and wing 
intensities in some wing bands. 
Since it is not possible to use the core intensity to obtain the  noise due to intrinsic changes, 
we use the average signal-to-noise of the left and right wings close to the calcium core 
as the signal-to-noise of the core. 
So the real signal-to-noise of the core is probably worse than that. 
The S/N ratio of the core and wing vary from 35 and 112 for two accumulations to 
174 and 216 for twenty accumulations.
We use the worse S/N ratio corresponding to an exposure time of 1.64 s and 
obtain the following values for the RMS errors of the core and wing intensities:
\begin{eqnarray} 
\frac{\sigma_c}{\textrm{I}_{\textrm{core}}} \sim \frac{1}{34} \sim 3\,\%\\
\frac{\sigma_w}{\textrm{I}_{\textrm{wing}}} \sim \frac{1}{112} \sim 1\,\%
\end{eqnarray} 

So, we can obtain the $\sigma_{c}$ as follows:
\begin{equation}
\frac{\sigma_x}{\textrm{x}} = 
\sqrt{(\frac{\sigma_c}{\textrm{I}_{\textrm{core}}})^2 + 
(\frac{\sigma_w}{\textrm{I}_{\textrm{wing}}})^2} \sim 3\,\%.
\end{equation} 

Error sources of the final calibrated data include uncertainties from  
the flat-field parameter, the straylight  and the linearity of the detector. 
The straylight uncertainty was obtained about 1\,\% in Sect. B.3. 
Assuming an upper limit for the uncertainty in the flat-field to be 5\,\%  
and some 1\,\% uncertainty in linearity of the core to wing ratio,  
{\textit{an upper limit for the uncertainty in the calibrated spectrum}} will be:

$$\frac{\sigma_{r}}{r} < 10\,\%.$$

\end{document}